\documentstyle[aps,prb,floats,graphicx,amssymb]{revtex}
\begin{document} \draft
\twocolumn[\hsize\textwidth\columnwidth\hsize\csname
@twocolumnfalse\endcsname 
\title{Periodic Vortex Structures in
Superfluid $^3$He-A} 
\author{J.M. Karim\"aki and E.V. Thuneberg} 
\address{Low Temperature Laboratory, Helsinki University of 
Technology, 02150 Espoo, Finland} 
\date{\today} 
\maketitle
\begin{abstract}  
We discuss the general properties of periodic vortex
arrangements in rotating superfluids. The different possible structures
are classified according to the symmetry space-groups and the
circulation number. We calculate numerically several types of vortex
structures in superfluid $^3$He-A. The calculations are done in the
Ginzburg-Landau region, but the method is applicable at all
temperatures. A phase diagram of vortices is constructed in the plane
formed by the magnetic field and the rotation velocity. The
characteristics of the six equilibrium vortex solutions are discussed.
One of these, the locked vortex 3, has not been considered in the
literature before. The vortex sheet forms the equilibrium state of
rotating $^3$He-A at rotation velocities exceeding 2.6 rad/s. The
results are in qualitative agreement with experiments.

\end{abstract} 
\bigskip ] 
\renewcommand{\topfraction}{1.0} 
\renewcommand{\textfraction}{0.0}

A superfluid cannot rotate homogeneously. Instead, quantized vortex
lines are present in the equilibrium rotating state of superfluid
$^4$He. In superfluid $^3$He the rotating states are more diverse. It
has been discussed recently by Parts {\it et.\ al.}\cite{diag} that
four different types of vortices have been found experimentally in the
superfluid A phase of $^3$He. In this paper we present theoretical
studies concerning the vortices observed in $^3$He-A. 

Some of the theoretical results that we presented in Refs.\
\onlinecite{diag,VSLT} were found to be incorrect in further 
calculations. These errors are corrected here. As a consequence, the
present phase diagram of vortices differs form the one in Refs.\
\onlinecite{diag,VSLT}. In particular, there appears a new vortex
structure, the locked vortex 3, but also the locations of other phase
boundaries are changed.  

For introduction to superfluid $^3$He \cite{LeggettRmp,VW} and its
vortices \cite{Fetterrev,SVrev,Volovikbook,Krusius,VSintro,VSLT} we
refer to various review articles. Although not introductory, this paper
intends to be a complete exposition of what is needed for understanding
the equilibrium vortex structures in bulk superfluid $^3$He-A.

We start in Section \ref{s.general} with the formulation of the vortex
problem, which is general enough for all superfluids and can be 
generalized also to superconductors. This gives a general
classification of vortex states based on space-group symmetry and
circulation number. The classification is continued in Section
\ref{s.avortex} using properties specific to $^3$He-A. The calculations
of the vortex structures are based on the hydrostatic theory, which is
discussed in Section \ref{s.hydro}, and the calculational method is
described in Section \ref{s.num}. Detailed description of the different
vortex types is given in Section \ref{s.results}. The correspondence
with experiments is discussed in Section \ref{s.exp}.

\section{The general vortex problem}\label{s.general}

Let us consider an uncharged fluid (in practice $^4$He or $^3$He) in a
container rotating at angular velocity $\bbox{\Omega}$. We will neglect
all complications arising from the finite size of the container.
Although we will not discuss the detailed correspondence, the analysis
in this section is also applicable to a charged fluid (superconductor)
when $\bbox{\Omega}$ is replaced by the averaged magnetic field ${\bf
B}$. 

At the microscopic level, the fluid has the effective Hamiltonian
$H_{\rm eff}=H_0-\bbox{\Omega}\cdot{\bf J}$. Here
$H_0=\sum_i(p_i^2/2m)+V$ is the Hamiltonian in a nonrotating system,
which consists of a kinetic energy term and an interaction energy
term $V$. The angular momentum ${\bf J}=\sum_i[{1\over 2}({\bf
r}_i\times{\bf p}_i-{\bf p}_i\times{\bf r}_i)+{\bf S}_i]$ consists of
an orbital and a spin part. We can write $H_{\rm eff}$ in the form 
\begin{eqnarray} H_{\rm eff}&=&\sum_i{1\over 2m}({\bf p}_i-m{\bf
v}_{{\rm n},i})^2+V\nonumber \\&&-\sum_i\bbox{\Omega}\cdot{\bf
S}_i-\sum_i{1\over 2}mv_{{\rm n},i}^2,  \label{e.hamfull}
\end{eqnarray} where ${\bf v}_{{\rm n},i}=\bbox{\Omega}\times{\bf r}_i$
is the ``normal fluid'' velocity at the location of the particle $i$.
The last term is the centrifugal energy. It causes the pressure to
increase with increasing distance from the rotation axis. In principle,
this term prohibits a strictly periodic vortex arrangement. However, it
is very small at experimentally relevant rotation velocities and
container sizes, so that we can safely neglect it. We will neglect also
the second-last term because it vanishes in $^4$He (S=0), and is very
small in $^3$He, where it corresponds to a magnetic field of $\approx
0.1\ \mu$T at a typical $\Omega=1$ rad/s. Because the rest of the paper
is based on the reduced $H_{\rm eff}$, we write it again: 
\begin{equation} H_{\rm eff}=\sum_i{1\over 2m}({\bf p}_i-m{\bf v}_{{\rm
n},i})^2+V. \label{e.ham} \end{equation}

We will classify the rotating states according to their symmetries.
For that purpose we first list all the symmetries of the Hamiltonian
(\ref{e.ham}). They are i) arbitrary translations, ii) arbitrary
rotations around $\bbox{\Omega}$ ($\infty_z$), iii) the combination
($2'_x$) of time inversion ($'$) and rotation by the angle $\pi$ (2)
around an axis perpendicular to $\bbox{\Omega}$, iv) the reflection
($m_z$) in the plane perpendicular to $\bbox{\Omega}$, v) the
combination ($m'_x$) of time inversion and reflection in a plane
containing $\bbox{\Omega}$, and vi) all combinations of these
operations. For each operation we have indicated a symbol in
parenthesis. They follow the international notation of crystallography
\cite{crystal} with a prime added to denote time inversion
\cite{magnetic}. Throughout this paper we use a rectangular coordinate
system $xyz$ where the $z$ axis is parallel to $\bbox{\Omega}$. 

It is not completely obvious that the translations perpendicular to
$\bbox{\Omega}$ are symmetry operations of $H_{\rm eff}$ (\ref{e.ham}).
This problem is equivalent to the case of electrons in uniform magnetic
field \cite{brown}, and the corresponding problem for superfluid order
parameter is discussed below. Another noteworthy feature is that both
$m_x$ and $2_x$ appear in combination with the time inversion.
Otherwise these operations would not preserve the direction of the
axial vector $\bbox{\Omega}$.

Generally, the physical system either has all the symmetries of the
Hamiltonian, or alternatively, one or more of the symmetries are
broken. An ordinary fluid would preserve all the symmetries of the
rotating Hamiltonian (\ref{e.ham}). This is not the case for a
superfluid. We will show below that at least part of the translation
symmetry is broken in the superfluid state when $\Omega\not= 0$. 

The fundamental property of superfluidity is that one quantum state
becomes macroscopically occupied. This condensate is described by an
order parameter $A({\bf r})$. The order parameter can be a scalar, as
in $^4$He, or a more complicated object. We associate a velocity field
${\bf v}_{\rm s}$ to the particles in the condensate. There is no
general expression for the superfluid velocity ${\bf v}_{\rm s}$ in
terms of $A({\bf r})$. Also, several different velocities ${\bf v}_{\rm
s}$ can exist, for example, one for spin up and another for spin down
particles. Irrespective of the precise definition, we only need to know
how ${\bf v}_{\rm s}$ changes in a gauge transformation. We require
that the velocity associated with the order parameter $\exp[{\rm
i}\phi({\bf r})]A^{(0)}({\bf r})$ is ${\bf v}_{\rm
s}=(\hbar/M)\bbox{\nabla}\phi({\bf r})+{\bf v}_{\rm s}^{(0)}$, where
${\bf v}_{\rm s}^{(0)}$ is the velocity corresponding to $A^{(0)}({\bf
r})$.  Here $M$ is a mass that depends on the particular system. It
equals the atomic mass for $^4$He ($M=m_4$) and twice the atomic mass
for $^3$He ($M=2m_3$).

It is now obvious that $A({\bf r})$ cannot be constant and also a
phase factor $\exp[{\rm i}\phi({\bf r})]$ times a constant is not
allowed. The reason is that the kinetic energy term in the Hamiltonian
(\ref{e.ham}) would grow faster than linearly with the volume of the
system because ${\bf v}_{\rm s}$ would be constrained by $\oint{\bf
v}_{\rm s}\cdot d{\bf r}=0$ and could not imitate ${\bf v}_{\rm
n}=\bbox{\Omega}\times{\bf r}$ on a large scale. 

Our basic assumption is that the equilibrium structure of the rotating
superfluid is periodic in space. It follows from above that the minimum
period has to be finite at least in one direction, which is not
parallel to $\bbox{\Omega}$. We do not make here any assumption whether
the translation symmetry is discrete or continuous in the two other
linearly independent directions. 

Crystalline materials are classified according to their symmetry into
1651 magnetic space groups. The most effective way to label these is the
international crystallographic notation \cite{crystal,magnetic}. We use
the same notation to label the space groups of rotating superfluids.
This is possible because corresponding to every rotating state there
exists at least one space group of a crystal. The reasons for this are
that i) the symmetries of the rotating-fluid problem listed above are a
subgroup of those possible for a crystal, and ii) no new symmetry
groups appear even if one or two of the translation symmetries in a
rotating superfluid were continuous. 

Not all the 1651 magnetic space groups are relevant for rotating fluids.
Firstly, the time inversion is present in a rotating superfluid in a
trivial way. In the generating symmetry operations it appears in
combinations with $m_x$ and $2_x$, and only with them. Thus, by simply
ignoring the time inversion, one can construct a one-to-one mapping
from symmetry operations of a rotating fluid into symmetry operations
that do not contain the time inversion \cite{tinote}. Thus it is
sufficient to limit to the 230 crystallographic space groups, which do
not include the time-inversion operation. Secondly, the number of
relevant groups is further reduced because a rotation axis higher than
2 is allowed in the direction of $\bbox{\Omega}$ only. This implies
that cubic groups are not acceptable. The remaining 194 space groups
each give rise to 1, 2, or 3 different symmetry groups of rotating
fluids. This is because some of the crystal groups can be oriented in
different ways relative to the $\bbox{\Omega}$ direction. 

What has been said above about symmetry does not directly apply to the
order parameter $A({\bf r})$. The reason is that this complex quantity
has phase $\phi$, which is not an observable quantity. Therefore,
instead of being strictly periodic, $A({\bf r})$ is only
quasiperiodic:   \begin{equation} A({\bf r}+{\bf a}_k) =\exp[{\rm
i}\phi_k({\bf r})]A({\bf r}).   \label{e.trans} \end{equation} Here
${\bf a}_k$ are three linearly independent translation vectors ($k=1$,
2, and 3) and $\phi_k({\bf r})$ are the corresponding phase shifts.
Similar phase factors occur also in rotations, reflections, and time
inversions. The quantity that has to be periodic in lattice
translations is ${\bf v}_{\rm s}-{\bf v}_{\rm n}$. This gives a
constraint for the phase shifts $\phi_k({\bf r})$. Using the gauge
invariance for ${\bf v}_{\rm s}$, one finds $\bbox{\nabla}\phi_k({\bf
r})= \tilde{\bbox{\Omega}}\times{\bf a}_k$. In order to simplify the
formulas, we will repeatedly use the notation
$\tilde{\bbox{\Omega}}=(M/\hbar)\bbox{\Omega}$. The gradient of
$\phi_k$ is trivially integrated to \cite{FNOT78} \begin{equation} 
\phi_k({\bf r})=C_k+ \tilde{\bbox{\Omega}}\times{\bf a}_k\cdot{\bf r}, 
\label{e.transphi} \end{equation}   where $C_k$ are constants of
integration. An implicit requirement here is that ${\bf v}_{\rm s}$ is
defined on the path of integration. We assume that the regions where 
${\bf v}_{\rm s}$ is undefined are at most one-dimensional. In this
case it seems possible to choose the unit cell of the translation
lattice so that ${\bf v}_{\rm s}$ is well defined along all its edges. 

An important requirement is that the lattice-translation rule
(\ref{e.trans}) is consistent with a uniquely defined $A({\bf r})$. We
express $A({\bf r}+{\bf a}_i+{\bf a}_j)$ as a function of $A({\bf r})$
using the translation rule twice. The result should be independent of
the order in which the two translations by ${\bf a}_i$ and ${\bf a}_j$
are done. This gives the condition  \begin{equation} 
\tilde{\bbox{\Omega}}\cdot{\bf a}_i\times{\bf a}_j =\pi
\sum_ke_{ijk}N_k,  \label{e.area3} \end{equation}  where $N_k$ are
integers and $e_{ijk}$ the fully antisymmetric tensor.

The lattice translation vectors ${\bf a}_k$ can be chosen in several
different ways. Next we want to redefine the set $\{{\bf a}_k\}$ so
that it is  optimal for further analysis. The new ${\bf a}_3$ can
always be chosen parallel to $\bbox{\Omega}$. Namely, setting ${\bf
a}_3=\sum_kN_k{\bf a}_k^{\rm old}$, it follows from Eq.\
(\ref{e.area3}) that ${\bf a}_3\times\bbox{\Omega}=0$. Applying Eq.\
(\ref{e.area3}) to a new linearly independent set $\{{\bf a}_k\}$, we
find that $N_1=N_2=0$ but $N_3\not = 0$. The nonzero integer value of
$N_3$ implies that a continuous translation symmetry can
exist only in the direction of $\bbox{\Omega}$. We can therefore
additionally require that ${\bf a}_1$ and ${\bf a}_2$ are primitive
translation vectors, {\it ie.}, they correspond to the minimum
(positive) value of $\bbox{\Omega}\cdot{\bf a}_1\times{\bf a}_2$.

We define the circulation number $N$ as equal to $N_3$ (\ref{e.area3})
corresponding to primitive ${\bf a}_1$ and ${\bf a}_2$:  
\begin{equation}  \tilde{\bbox{\Omega}}\cdot{\bf a}_1\times{\bf a}_2
=\pi N.  \label{e.area}\end{equation}   Similar to the symmetry groups,
the different values of $N$ can be used to classify the rotating
states. $N$ is called the circulation number because it is related to
the circulation of the superfluid velocity around a primitive cell 
\begin{equation} 
 N={1\over 2\pi}\oint_{\rm primitive\ cell}d{\bf r}\cdot
\tilde{\bf v}_{\rm s}, 
\label{e.circ} \end{equation}   
where $\tilde{\bf v}_{\rm s}=(M/\hbar){\bf v}_{\rm s}$. We note that the
limitation to the boundary of the primitive cell in Eq.\ (\ref{e.circ})
is crucial in $^3$He-A, where the circulation is not generally
quantized.

Let us consider the case that there is a continuous translation
symmetry along $\bbox{\Omega}$. This is an important case because all
known vortex types belong to this category. However, very few general
properties can be listed in addition to those ones already mentioned
above. The main simplification is that the primitive translation
vectors ${\bf a}_1$ and ${\bf a}_2$ can be chosen perpendicular to
$\bbox{\Omega}$. These generate a two dimensional Bravais lattice. Thus
these rotating states can be classified into five categories
according to the symmetry of the 2-D lattice \cite{crystal}: oblique,
square, hexagonal, primitive rectangular and centered rectangular. The
number of possible space groups is considerably larger. In particular,
the 17 two dimensional space groups listed in Ref. \onlinecite{crystal}
are not sufficient for rotating states because they lack the
operation $m_z$.

We illustrate the classification of vortices with known example cases.
For a scalar order parameter ($^4$He) the Bravais lattice is hexagonal
and $N=1$ \cite{Tka}. It has the symmetry group $P{6\over m}{2'\over
m'}{2'\over m'}$, or shortly $P6/mm'm'$. Generally, the international
symbols consist of a letter followed by three symbol sets
\cite{crystal}. The letter shows the basis of the lattice, for example,
$P$ denotes a primitive and $C$ a centered unit cell. The following
three symbol sets describe symmetries with respect to three different
inequivalent axes, respectively. The first ${6\over m}$ or ${6/m}$
indicates that there is a 6-fold rotation symmetry and a reflection
symmetry $m$, both with respect to the same axis. Here the 6 fold axis
has to be parallel to $\bbox{\Omega}$ and thus the reflection plane is
perpendicular to $\bbox{\Omega}$. The second set ${2'\over m'}$
describes a $2'$ symmetry and an $m'$ symmetry with respect to an axis
perpendicular to $\bbox{\Omega}$. Finally, the third set ${2'\over m'}$
describes the same symmetries around the third inequivalent axis of
the hexagonal lattice. 

The relative orientation of the space group and $\bbox{\Omega}$ is
usually revealed by the primes because the primed axes are always
perpendicular to $\bbox{\Omega}$. For some structures ($C12'1$, for
example) pure symmetry considerations are insufficient to fix the
direction of $\bbox{\Omega}$. However, as proved below Eq.\
(\ref{e.area3}), the $\bbox{\Omega}$ axis always coincides with one
direction of translation symmetry. This is not a consequence of
symmetry but arises from the divergent rigidity of the vortex lattice
against tilt deformation at long wave lengths \cite{Sonin}.

As another example, we consider vortices of $^3$He-B. Two types of fully
stable vortices are known. An isolated A-phase core vortex has symmetry
class $\infty m'$, and a double-core vortex $2m'm'$ \cite{T87}. Here
the symmetry breaking relative to ${\infty\over m}{2'\over m'}$ arises
from the core of each vortex. When these vortices form a lattice, the
simplest possible structures have $N=1$, and the space groups are
$P6m'm'$ and $Cm'm'2$, respectively. The lattice breaks the rotating
symmetry of the A-phase-core vortex to six-fold, and the two-fold
rotation symmetry of the double-core vortex breaks the hexagonal
lattice symmetry to centered rectangular. Both these effects are in
practice very weak because the core sizes of the vortices in $^3$He-B
are much smaller than the distance between vortices.  

The symmetry classification of vortices has previously been made only
for point groups. This means that all the translations in the space
group are ignored. Although this does not describe the whole lattice
symmetry, there are several physical properties for which the point
group gives a sufficient description \cite{SVrev,SF85}. We also comment
on the notation. The symmetries $\bar 1$, $m'_y$, $2'_y$  often appear
in dealing with vortices of $^3$He. Here $\bar 1$ is the inversion and
the sub-indexes in $m'_y$ and $2'_y$ denote that they refer to the same
axis. These operations give rise to five symmetry classes $1$, $\bar
1$, $m'$, $2'$, and ${2'\over m'}$. Here the first one contains only
the unit element, the three middle ones have each one symmetry
operation $\bar 1$, $m'_y$, or $2'_y$, respectively, and the last one
has all the three (because two of them imply the third). In Ref.\
\onlinecite{SVrev}, the same groups were labeled by letters $o$, $u$,
$v$, $w$ and $uvw$, respectively. Still another notation is due to
Sch\"onflies, and this was used to denote the same classes in Ref.\
\onlinecite{SF85}. Contrary to the international crystallographic
symbols, these other notations do not allow a meaningful generalization
to space groups. 

Let us study the meaning of the constants $C_k$ in Eq.\
(\ref{e.transphi}). We will show that $C_1$ and $C_2$ can be put to
zero without losing generality. We consider an arbitrary order
parameter field $A^{(0)}({\bf r})$. We construct from it another field
$A({\bf r})$ by doing a translation by an arbitrary vector ${\bf b}$ as
follows:  \begin{equation}  A({\bf r})=\exp({\rm
i}\tilde{\bbox{\Omega}}\times{\bf b}\cdot{\bf r})A^{(0)}({\bf r}-{\bf
b}).  \label{e.shift} \end{equation}   This field obviously has the
same energy as the original one because the phase factor takes care
that the counterflow velocity ${\bf v}={\bf v}_{\rm s}-{\bf v}_{\rm n}$
is unchanged: ${\bf v}({\bf r})={\bf v}^{(0)}({\bf r}-{\bf b})$. By
straightforward calculation one can verify that the coefficients $C_1$
and $C_2$ for the new field are related to the old ones by
$C_k=C_k^{(0)}+2\tilde{\bbox{\Omega}}\times{\bf b}\cdot{\bf a}_k$.
Choosing ${\bf b}$ appropriately, one can put $C_1$ and $C_2$ to zero.
Thus the significance of $C_1$ and $C_2$ is that their values fix the
position of the vortex solution relative to the rotation center.

The coefficient $C_3$ in Eq.\ (\ref{e.transphi}) is the phase shift in
translations parallel to ${\bbox{\Omega}}$. It is related to the
superfluid velocity parallel to ${\bbox{\Omega}}$. It often vanishes
for symmetry reasons, but it can be nonzero for vortices of low
symmetry. For example, consider a vortex with the symmetry group
$C12'1$ and a continuous translation symmetry in the $z$ direction. The
only point symmetry operation $2'_x$ leaves the $z$ component of the
current unchanged. Thus, such a vortex generally has a nonzero net
superfluid current in the $z$ direction even though $v_{{\rm s},z}=0$.
Depending on the boundary conditions at $z=\pm\infty$, this current may
be compensated by a current arising from a nonzero $\tilde v_{{\rm
s},z}=C_3/a_3$.

\section{Superfluid $^3$H\lowercase{e}-A}\label{s.avortex}

The previous section showed that the vortex structures in any
superfluid can be classified according to the circulation number and
the space group. In this section we continue the classification using
properties specific to $^3$He-A. 

The order parameter of bulk superfluid $^3$He-A is a complex $3\times 3$
matrix of the form \cite{LeggettRmp,VW}   
 \begin{equation} \tensor{A}=\Delta\hat{\bf d}(\hat{\bf m}+{\rm
i}\hat{\bf n})\ .  \label{e.op} \end{equation}  
 Here $\hat{\bf d}$, $\hat{\bf m}$ and $\hat{\bf n}$ are unit vectors
and $\hat{\bf m}\perp \hat{\bf n}$. The amplitude $\Delta$ is a
temperature and pressure dependent constant. It is conventional to
define $\hat{\bf l}=\hat{\bf m}\times\hat{\bf n}$, so that $\hat{\bf
m}$, $\hat{\bf n}$ and $\hat{\bf l}$ form an orthonormal set. 

As a first step, the vortices are classified to ``continuous'' and
``singular''. The former alternative means that the bulk form
(\ref{e.op}) with constant $\Delta$ forms a good approximation to the
order parameter everywhere in the primitive cell. The latter
alternative means that this is not the case. This classification may
not be precise in general, but there is no difficulty for the six
vortex types to be considered here: only the ``singular vortex'' is
singular, the other four are continuous. 

We note that only singular vortices exist for a scalar $A$ because the
amplitude of $A$ has to vanish somewhere within the primitive cell of
$N=1$. Continuous vortices are possible in $^3$He-A because nonzero
circulation can be generated by appropriate $\hat{\bf m}({\bf r})$ and
$\hat{\bf n}({\bf r})$ fields.

The continuous structures can be further  classified by the numbers
\begin{eqnarray}  \nu_d&=&{1\over 4\pi} \int_{\rm primitive\ cell}
dx\,dy~\hat {\bf d} \cdot{\partial\hat {\bf d}\over\partial x} \times
{\partial \hat {\bf d}\over\partial y} \label{e.nud}\\ \nu_l&=&{1\over
4\pi} \int_{\rm primitive\ cell} dx\,dy~\hat {\bf l} \cdot{\partial\hat
{\bf l}\over\partial x} \times {\partial \hat {\bf l}\over\partial y}. 
\label{e.nul} \end{eqnarray}  
 These numbers are integers because $\hat{\bf d}$ and $\hat{\bf l}$ are
periodic. They describe how many times the mapping from the primitive
cell to the vectors $\hat{\bf d}$ and
$\hat{\bf l}$ covers the unit spheres.  

The numbers $N$ and $\nu_l$ are not independent. This follows from the
definition of the superfluid velocity  \begin{equation} \tilde{\bf
v}_{\rm s}=\sum_im_i\bbox{\nabla} n_i.  \label{e.vel} \end{equation} 
 (As above, we use $\tilde{\bf v}=(2m_3/\hbar){\bf v}$, where $m_3$ is
the mass of a $^3$He atom.) This implies the Mermin-Ho relation
\cite{MerminHo}  \begin{equation} \bbox{\nabla}\times\tilde{\bf v}_{\rm
s} ={1\over 2}\sum_{ijk}e_{ijk}l_i\bbox{\nabla} l_j\times\bbox{\nabla}
l_k,  \label{e.mh} \end{equation} 
 which together with Eqs.\ (\ref{e.circ}) and (\ref{e.nul}) gives 
\begin{equation} N=2\nu_l. \label{e.contn} \end{equation}

\section{Hydrostatic theory}\label{s.hydro}

For a quantitative determination of the vortex structures we use an
energy functional $F(A)$. In principle, it can be calculated from the
effective Hamiltonian (\ref{e.ham}) as $F(A)=-T\ln[\mathop{\rm
Tr}\exp(-H_{\rm eff}/T)]$. This is a functional of $A({\bf r})$
because the trace ($\mathop{\rm Tr}$) is restricted to states having a
given macroscopic $A({\bf r})$. Various approximations for $F(A)$ are
available: quasiclassical weak-coupling and weak-coupling-plus models,
and phenomenological theories such as the Ginzburg-Landau theory and
the hydrodynamic theory.

The basic assumption of the hydrodynamic theory is that the deviation
of the order parameter $\tensor{A}$ from the bulk form (\ref{e.op}) is
small. For this we have to require two conditions. (i) The magnetic
field ${\bf H}$ should not be too large. In practice this condition
excludes only a small region near the superfluid transition temperature
$T_{\rm c}$, where the A phase is distorted towards the A$_1$ phase
\cite{VW}. (ii) The vectors $\hat{\bf d}({\bf r})$, $\hat{\bf m}({\bf
r})$ and $\hat{\bf n}({\bf r})$ are sufficiently slowly varying
functions of the location ${\bf r}$. This implies that the hydrodynamic
approach can be used for continuous vortices, but it is insufficient
for singular ones. 

Because of the slow variation, only terms up to the second order in the
gradients of $\hat{\bf d}$, $\hat{\bf m}$ and $\hat{\bf n}$ are needed
in the energy functional. The functional can be written as 
\begin{equation} F={1\over V}\int_V d^3r(f_{\rm d}+f_{\rm h}+f_{\rm
g}). \label{e.fstatic} \end{equation} Here the volume $V$ of
integration is assumed to consist of an (arbitrary) integral number of
unit cells.  The magnetic dipole-dipole interaction $f_{\rm d}$ is
given by \cite{Leggett}   
\begin{equation} f_{\rm d}={1\over 2}g_{\rm
d}\vert\hat{\bf d}\times\hat{\bf l}\vert^2. 
\label{e.fd} \end{equation}
 The magnetic anisotropy term is  
\begin{equation} f_{\rm h}={1\over
2}g_{\rm h}(\hat{\bf d}\cdot{\bf H})^2, 
\label{e.fh} \end{equation}
 and the gradient energy \cite{Cross}  
\begin{eqnarray} 2f_{\rm g}&=&
\rho_\perp{\bf v}^2+(\rho_\parallel-\rho_\perp)(\hat{\bf l}\cdot{\bf
v})^2 \nonumber \\&&+2C{\bf v}\cdot\bbox{\nabla}\times\hat{\bf l}
-2C_0(\hat{\bf l}\cdot{\bf v}) (\hat{\bf
l}\cdot\bbox{\nabla}\times\hat{\bf l})\nonumber \\&&+ K_{\rm
s}(\bbox{\nabla}\cdot\hat{\bf l})^2 +K_{\rm t}(\hat{\bf
l}\cdot\bbox{\nabla}\times\hat{\bf l})^2\nonumber \\ &&+K_{\rm
b}\vert\hat{\bf l}\times(\bbox{\nabla}\times\hat{\bf l})\vert^2 +K_5
\vert(\hat{\bf l}\cdot\bbox{\nabla})\hat{\bf d}\vert^2\nonumber \\&&+
K_6\sum_{ij}[(\hat{\bf l}\times\bbox{\nabla})_i\hat{\bf d}_j)]^2.
\label{e.grad} \end{eqnarray} 
 The gradient term includes also the
kinetic energy, which is a function of the counterflow velocity ${\bf
v}={\bf v}_{\rm s}-{\bf v}_{\rm n}$. It follows from the structure of
the functional that $F_{\rm eq}(H,\Omega)$ of the equilibrium state is
a non-decreasing function of both $H$ and $\Omega$.  The energy is
normalized so that $F_{\rm eq}(H,0)=0$.  

It should be noted that ${\bf v}_{\rm s}$ and $\hat{\bf l}=\hat{\bf
m}\times\hat{\bf n}$ are not completely independent variables but are
constrained by the Mermin-Ho relation (\ref{e.mh}). In order to avoid
such complicated constraints, we use $\hat{\bf d}$, $\hat{\bf m}$ and
$\hat{\bf n}$ as the basic variables. With these variables the
constraints are simpler: $\hat{\bf d}$, $\hat{\bf m}$ and $\hat{\bf n}$
have to be unit vectors and $\hat{\bf m}\perp\hat{\bf n}$. The energy
functional (\ref{e.fstatic}) can be expressed as a function of
$\hat{\bf d}$, $\hat{\bf m}$ and $\hat{\bf n}$. Only the gradient terms
require some calculation, and we get  
\begin{eqnarray} 2f_{\rm g}&=&
(\bar\rho_\parallel+2K_7)\sum_i(\hat{\bf m}\cdot D_i\hat{\bf n})^2 
\nonumber \\& +&(K_{\rm s}+K_7)\sum_i[(\hat{\bf m}\cdot{\bf D}\hat
m_i)^2+(\hat{\bf n}\cdot{\bf D}\hat n_i)^2] \nonumber \\& +&2K_{\rm
s}\sum_i(\hat{\bf m}\cdot{\bf D}\hat m_i)\hat{\bf n}\cdot{\bf D}\hat
n_i \nonumber \\& +&(K_{\rm t}+K_7)\sum_i[(\hat{\bf n}\cdot{\bf D}\hat
m_i)^2+(\hat{\bf m}\cdot{\bf D}\hat n_i)^2] \nonumber \\& -&2K_{\rm
t}\sum_i(\hat{\bf n}\cdot{\bf D}\hat m_i)\hat{\bf m}\cdot{\bf D}\hat n_i
\nonumber \\& +&(K_{\rm b}+\bar C-\bar C_0+K_7)[({\bf D}\cdot\hat{\bf
m})^2+({\bf D}\cdot\hat{\bf n})^2] \nonumber \\& +&2(K_{\rm b}-\bar
C_0+K_7)\sum_i[({\bf D}\cdot\hat{\bf m})\hat m_i\hat{\bf n}\cdot{\bf
D}\hat n_i \nonumber \\&&
   -({\bf D}\cdot\hat{\bf n})\hat m_i\hat{\bf m}\cdot{\bf D}\hat n_i]
\nonumber \\& +&(\bar C_0-\bar C)\sum_{ik}[(D_i\hat m_k)D_k\hat
m_i+(D_i\hat n_k)D_k\hat n_i] \nonumber \\& -&K_7\sum_{ik}[(D_i\hat
m_k)^2+(D_i\hat n_k)^2] \nonumber \\& +&K_5\sum_{ik}(\nabla_i\hat d_k)^2
\nonumber \\& +&(K_6-K_5)\sum_i[(\hat{\bf m}\cdot\bbox{\nabla}\hat
d_i)^2 +(\hat{\bf n}\cdot\bbox{\nabla}\hat d_i)^2]. \label{e.grad2}
\end{eqnarray} 
 Here we use gauge invariant derivatives ${\bf D}\hat
m_i=\bbox{\nabla}\hat m_i+\tilde{\bf  v}_{\rm n}\hat n_i$ and ${\bf
D}\hat n_i=\bbox{\nabla}\hat n_i-\tilde{\bf  v}_{\rm n}\hat m_i$. We
have also used the notations $\tilde{\bf v}_{\rm
n}=(2m_3/\hbar)\bbox{\Omega}\times{\bf r}$,
$\bar\rho=(\hbar/2m_3)^2\rho$, $\bar C=(\hbar/2m_3)C$, and 
$K_7=\bar\rho_\perp-\bar\rho_\parallel-K_{\rm s}-K_{\rm t}+K_{\rm
b}-2\bar C_0$.

One can transform the gradient energy (\ref{e.grad2}) by partial
integration. For example, $\int d^3r(\nabla_i\hat m_k)\nabla_k\hat
m_i=\int d^3r(\bbox{\nabla}\cdot\hat{\bf m})^2$ plus a surface term.
For the present purposes such partial integrations can be done
without paying attention to the surface terms. The reasons are that i)
the surface terms can affect the equilibrium configuration only near
surfaces, if anywhere, and ii) although the local energy density is
changed in the partial integration, all the energies of vortices are
unchanged because of the periodic boundary conditions. 

The full gradient energy (\ref{e.grad2}) is written down here in order
to demonstrate that our calculational method is feasible whenever the
hydrodynamic approximation is valid. In particular, the theory applies
to all temperatures $T<T_{\rm c}$ except a small region near $T_{\rm
c}$ (due to the A$_1$ phase) and another region around $T=0$. However,
the present numerical calculations are made in the Ginzburg-Landau (GL)
region \cite{T87}. This means temperatures only near $T_{\rm c}$
($T_{\rm c}-T\ll T_{\rm c}$), but this range is still wider than the
one that has to be excluded because of distortion towards the A$_1$
phase. In this region the GL theory is more general than the
hydrodynamic one. If one makes the hydrodynamic approximation in the GL
theory, one arrives at the set of equations presented above, but with
certain restrictions on the coefficients. They are
$\bar\rho_\parallel/2=\bar\rho_\perp/(\gamma+1)=\bar
C/(\gamma-\eta-1)=\bar C_0/(\gamma-1)=K_{\rm s}=K_{\rm t}=K_{\rm
b}/\gamma=K_5/2=K_6/(\gamma+1)$. These conditions imply that all terms
that are higher than second order in $\hat{\bf m}$ or $\hat{\bf n}$
disappear from the gradient energy (\ref{e.grad2}).

Collecting all simplifications and reducing units, we can write the
energy terms as   \begin{eqnarray} f_{\rm d}&=&{1\over 2}[(\hat{\bf
d}\cdot\hat{\bf m})^2+(\hat{\bf d} \cdot\hat{\bf n})^2]\nonumber\\
f_{\rm h}&=&{1\over 2}(\hat{\bf d}\cdot{\bf H})^2\nonumber \\ 4f_{\rm
g}&=&\sum_{ik}[\nabla_i\hat m_k+(v_{\rm n})_in_k]^2
+\sum_{ik}[\nabla_i\hat n_k-(v_{\rm n})_im_k]^2 \nonumber \\&&
+(\gamma-1)[(\bbox{\nabla}\cdot\hat{\bf m} +{\bf v}_{\rm
n}\cdot\hat{\bf n})^2 +(\bbox{\nabla}\cdot\hat{\bf n} -{\bf v}_{\rm
n}\cdot\hat{\bf m})^2 \nonumber \\&&+\sum_i(\hat{\bf
m}\cdot\bbox{\nabla}\hat d_i)^2 +\sum_i(\hat{\bf
n}\cdot\bbox{\nabla}\hat d_i)^2] \nonumber \\&&
+2\sum_{ik}(\nabla_i\hat d_k)^2+4(2\eta-1){\bf m}\times{\bf
n}\cdot\bbox{\Omega}, \label{e.gle} \end{eqnarray} where ${\bf v}_{\rm
n}=\bbox{\Omega}\times{\bf r}$. We have used ``dipole units'' for
length ($\xi_{\rm d}=(\hbar/2m_3)\sqrt{\rho_\parallel/g_{\rm d}}$),
field ($H_{\rm d}=\sqrt{g_{\rm d}/g_{\rm h}}$), angular velocity
($\Omega_{\rm d}=\hbar/2m_3\xi_{\rm d}^2$), and energy density ($g_{\rm
d}$). 

It should be noted that the functional (\ref{e.gle}) is based on purely
phenomenological considerations. These leave open only two dimensionless
parameters $\gamma$ and $\eta$. (In the notation of Ref.\
\onlinecite{SRres}, $\gamma=K_L/K_T$ and $\eta=K_C/K_T$.) They can be
calculated in the quasiclassical theory. In the weak coupling
approximation $\gamma=3$ and $\eta=1$. A more complicated
weak-coupling-plus model gives $\gamma\approx 3.1$, but $\eta$ is
unchanged \cite{SRres}. We have made a few tests that our results are
not sensitive to the value of $\gamma$, so we will use the
weak-coupling values in the following.

\section{Numerical calculation}\label{s.num}

For numerical computation the energy functional (\ref{e.gle}) was
discretized on a lattice. We assume that there is no dependence on $z$,
so that a two-dimensional square lattice $(x_j, y_k)$ is sufficient.
The first thing in the numerical program is to specify the magnetic
field, the rotation velocity $\Omega$, the circulation number $N$ and
the Bravais lattice. For rectangular lattices one needs to specify the
ratio $b/a$, where $a$ denotes the length of the shortest possible
primitive vector, and $b$ is the lattice constant in the perpendicular
direction in the rectangular cell. No oblique lattices were found. The
area of the primitive cell was then determined from Eq.\
(\ref{e.area}). For simplicity of boundary conditions, it was useful to
choose the lattice constant of the calculational lattice commensurate
with the primitive cell of the vortex lattice. The last preparatory
step was to give an initial guess for the fields $\hat{\bf d}$,
$\hat{\bf m}$ and $\hat{\bf n}$.  Then the values $\hat{\bf d}(x_j,
y_k)$, $\hat{\bf m}(x_j, y_k)$ and $\hat{\bf n}(x_j, y_k)$ were
changed iteratively. At each lattice point $(x_j, y_k)$ in the
primitive cell, a torque acting on the spin  vector ($\hat{\bf d}$) and
on the orbital vectors ($\hat{\bf m}$ and $\hat{\bf n}$) was
calculated, and the vectors were rotated proportionally to the torque
\cite{TK94}. The proportionality coefficient was chosen experimentally
to achieve fast convergence. The values of the fields outside the
primitive cell were determined from the periodic boundary conditions
(\ref{e.trans}-\ref{e.transphi}). As discussed above, the coefficients
$C_1$ and $C_2$ are arbitrary but it was natural to choose them
consistent with the initial guess in order to avoid unnecessary motion
of the vortex. We assume $C_3=0$. The iteration was continued until the
energy converged, and the torques approached zero. 

For rectangular lattices, the energy $F$ needs to be minimized also
with respect to the ratio of the lattice constants $u=b/a$. This
process is considerably simplified by noting that a calculation at a
given value of $u$ not only gives $F(u)$, but also the first derivative
$dF(u)/du$ at constant area $ab$. It follows from the stationarity of
the energy functional that  \begin{eqnarray} {dF(u)\over du}&=&{1\over
2uV}\int_Vd^3r\sum_{ij} \bigg({\partial f_{\rm g}\over\partial
D_xA_{ij}}D_xA_{ij} \nonumber\\&& -{\partial f_{\rm g}\over \partial
D_yA_{ij}}D_yA_{ij}+ {\rm c.c.}\bigg).
\label{e.latticestrecth}\end{eqnarray} Here $A_{ij}$ is the general
order-parameter matrix in superfluid $^3$He and c.c.\ the complex
conjugate. Application to functional (\ref{e.gle}) gives
\begin{equation} {dF(u)\over du}={1\over uV}\int_Vd^3r
\left(T_x-T_y\right), \label{e.glresult}\end{equation} where
\begin{eqnarray} 4T_i&=&\sum_k(\nabla_i\hat m_k+v_{{\rm n}i}\hat n_k)^2
+\sum_k(\nabla_i\hat n_k-v_{{\rm n}i}\hat m_k)^2 \nonumber\\&&
+(\gamma-1)\bigg[(\nabla_i\hat m_i +v_{{\rm n}i}\hat n_i)
(\bbox{\nabla}\cdot\hat{\bf m} +{\bf v}_{\rm n}\cdot\hat{\bf n})
\nonumber\\&& +(\nabla_i\hat n_i-v_{{\rm n}i}\hat m_i)
(\bbox{\nabla}\cdot\hat{\bf n} -{\bf v}_{\rm n}\cdot\hat{\bf m})
\nonumber\\&& +\sum_k(\hat m_i\nabla_i\hat d_k)\hat{\bf
m}\cdot\bbox{\nabla}\hat d_k +\sum_k(\hat n_i\nabla_i\hat d_k)\hat{\bf
n}\cdot\bbox{\nabla}\hat d_k \bigg] \nonumber\\&& +2\sum_k(\nabla_i\hat
d_k)^2. \label{e.glpart}\end{eqnarray}

The iteration was often started with a rather small number of lattice
points ($\sim 1000$), and this was iterated to convergence. Then new
lattice points were added in between the old ones, and the iteration
was continued. The maximum final lattices contained around 100 000
points.

The procedure was repeated for several different initial guesses and
values of magnetic field and rotation velocity.  Although the
finding of the minimum-energy structures is a well defined mathematical
problem, physical intuition is needed in inventing the initial guesses.
Particularly good guesses are the models used in previous
investigations of vortices. In addition, we tried several variants of
these. The initial guesses leading to the different structures are
listed in the Appendix.

\section{Results}\label{s.results}

\begin{figure}[bt]
\begin{center}\leavevmode
\includegraphics[width=0.9\linewidth]{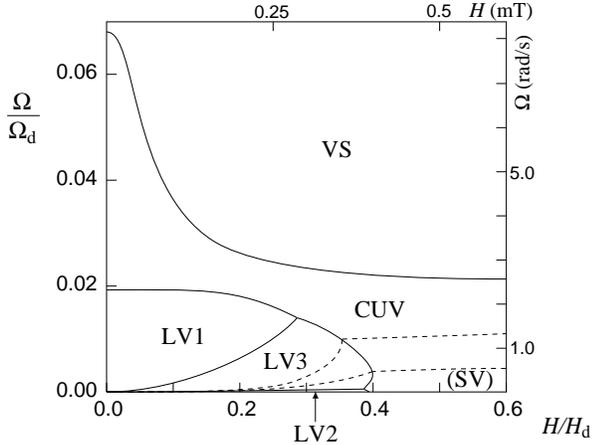}
\bigskip
\caption{ 
The phase diagram of vortices in the plane formed
by the magnetic field ${\bf H}$ and the rotation velocity
$\protect\bbox{\Omega}\parallel{\bf H}$. The calculation is done using
the dimensionless units given below Eq.\ (\protect\ref{e.gle}). The
solid lines denote phase boundaries between different continuous
vortices (LV1, LV3, CUV, and VS). The LV2 should become the equilibrium
locked vortex at very low $\Omega\lesssim 0.0005\Omega_{\rm d}$.
The dashed lines denote the phase boundaries of the singular vortex
(SV) against the LV3 and the CUV. These are calculated using two values
of the SV energy parameter: $c=2.6$ (upper dashed line) and $c=3.1$
(lower dashed line). The real units for $\Omega$ and $H$ are calculated
using $\Omega_{\rm d}=120$ rad/s and $H_{\rm d}=2.0$ mT, which are
estimated for the pressure of 29 bar.
}\label{f.diagram}\end{center}\end{figure}

In all calculations $T\approx T_{\rm c}$ and the field ${\bf H}$ was
chosen parallel to the rotation axis $z$. The phase diagram of vortices
in the plane formed by the magnetic field $H$ and the rotation velocity
$\Omega$ is shown in Fig.\ \ref{f.diagram}. There are six equilibrium
vortex types, which are discussed separately below. In the names of the
vortices, we follow Ref.\ \onlinecite{diag}. The energy densities $F$
(\ref{e.fstatic}) of the vortices are expressed in reduced units
defined under Eq.\ (\ref{e.gle}). We choose the $x$ axis parallel to
the shortest primitive vector of the two-dimensional Bravais lattice.

(i) The locked vortex 1 (LV1) has the quantum numbers $N=4$ and
$\nu_l=\nu_d=2$ \cite{FNOT78,NOTF79,FSS83,F85,ZM85,TPM92}. Its most
distinguishing feature is the square Bravais lattice. The space group
is $P{4\over n}{2'\over b'}{2'\over m'}$, or shortly $P4/nb'm'$. The
symmetry operations of this as well as other space groups are listed in
Ref.\ \onlinecite{crystal}. 
\begin{figure}[bt]
\begin{center}\leavevmode
\includegraphics[width=0.8\linewidth]{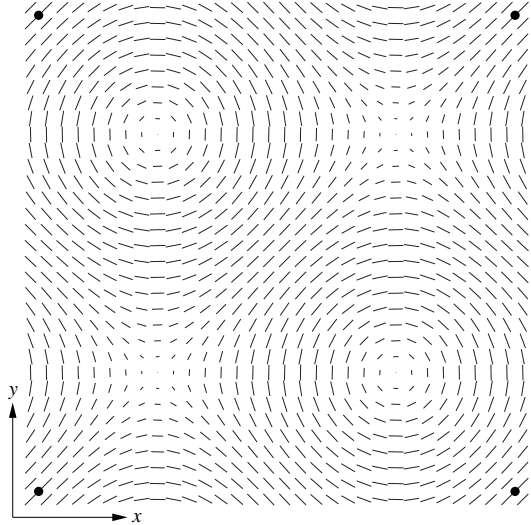}
\bigskip
\caption{ 
Locked vortex 1   (LV1) in zero field. In figures
\protect\ref{f.LV1}--\protect\ref{f.VS}, the short lines denote the
projection of the unit vector $\hat{\bf l}$ to the $x-y$ plane. 
For clarity of figure, the heads of the arrows are omitted. The
component $\hat l_z$  is positive ($\hat{\bf
l}\cdot\protect\bbox{\Omega}>0$) in regions where $(\hat l_x, \hat
l_y)$ has circular appearance, and negative in regions where $(\hat
l_x, \hat l_y)$ is hyperbolic. The dots denote equivalent points in
the  periodic lattice. The LV1 has the square Bravais lattice and the
space group $P4/nb'm'$. 
}\label{f.LV1}\end{center}\end{figure}

Qualitatively the structure can be understood so that the first thing to
minimize is the dipole energy (\ref{e.fd}). This gives a ``locked''
configuration where $\hat {\bf d}=\hat {\bf l}$ everywhere. If the
field is zero, then the energy of this structure arises solely from the
gradient terms (\ref{e.grad}). These are minimized by a smooth
distribution where the gradient of $\hat {\bf l}$ has the same order
of magnitude everywhere. Numerical calculations show that this is
achieved in a square lattice with $N=4$, see Fig.\ \ref{f.LV1}. 

The structure can be interpreted to consist of four elementary units.
These units are called Mermin-Ho vortices because of a resembling
structure first described in a cylindrical container \cite{MerminHo}.
The boundary of each unit can be defined by $\hat l_z=0$. Two of the
four units have a circular distribution of $\hat{\bf l}$ and $\hat
l_z>0$ ($\hat{\bf l}\cdot\protect\bbox{\Omega}>0$). The other two have
a hyperbolic distribution and $\hat l_z<0$. It follows from Eq.\
(\ref{e.mh}) that each elementary vortex contributes a unity to $N=4$.
A finite axial field makes the cores of the Mermin-Ho vortices
shrink because the field favors the orientation $\hat{\bf
d}\perp\hat{\bf z}$ at the borders of the elementary vortices. 

At $H=0$ we find the energy $F=4.72\Omega^{0.997}$ in the range
$\Omega = 0.005\ldots 0.04$. In the case
of perfect locking ($\hat {\bf d}\equiv\hat {\bf l}$), the energy would
be strictly linear in $\Omega$. We find that this is only an
approximation because with increasing $\Omega$ the kinetic energy
(\ref{e.grad}) is reduced at the expense of the dipole energy
(\ref{e.fd}). Our energy can be compared with Ref.\
\onlinecite{FNOT78}, where a variational calculation gives a 30\%  and a
model based on isolated Mermin-Ho vortices a 6\% larger value. 

A possible competitor to the LV1 structure is a hexagonal lattice with
$N=2\nu_l=2\nu_d=2$\cite{VK77}. The ansatz form has $\hat{\bf
l}\parallel\bbox{\Omega}$ at the borders of the Wigner-Seitz cell.  The
space group is either $P6m'm'$ for a radial and $P62'2'$ for a circular
distribution of $\hat{\bf l}$. In agreement with previous authors, we
find that this structure does not correspond to the minimum energy at
any values of $\Omega$ and $H$ \cite{FNOT78,FSS83,TPM92}. 

(ii) The locked vortex 3 (LV3) has the same quantum numbers $N=4$ and
$\nu_l=\nu_d=2$ as LV1. The main difference is that the lattice
structure is primitive rectangular rather than a square.  The space
group is $P{2'\over b'}{2'\over a'}{2\over n}$, or shortly $Pb'a'n$.
This structure has not been previously studied in the literature.

\begin{figure}[bt]
\begin{center}\leavevmode
\includegraphics[width=0.75\linewidth]{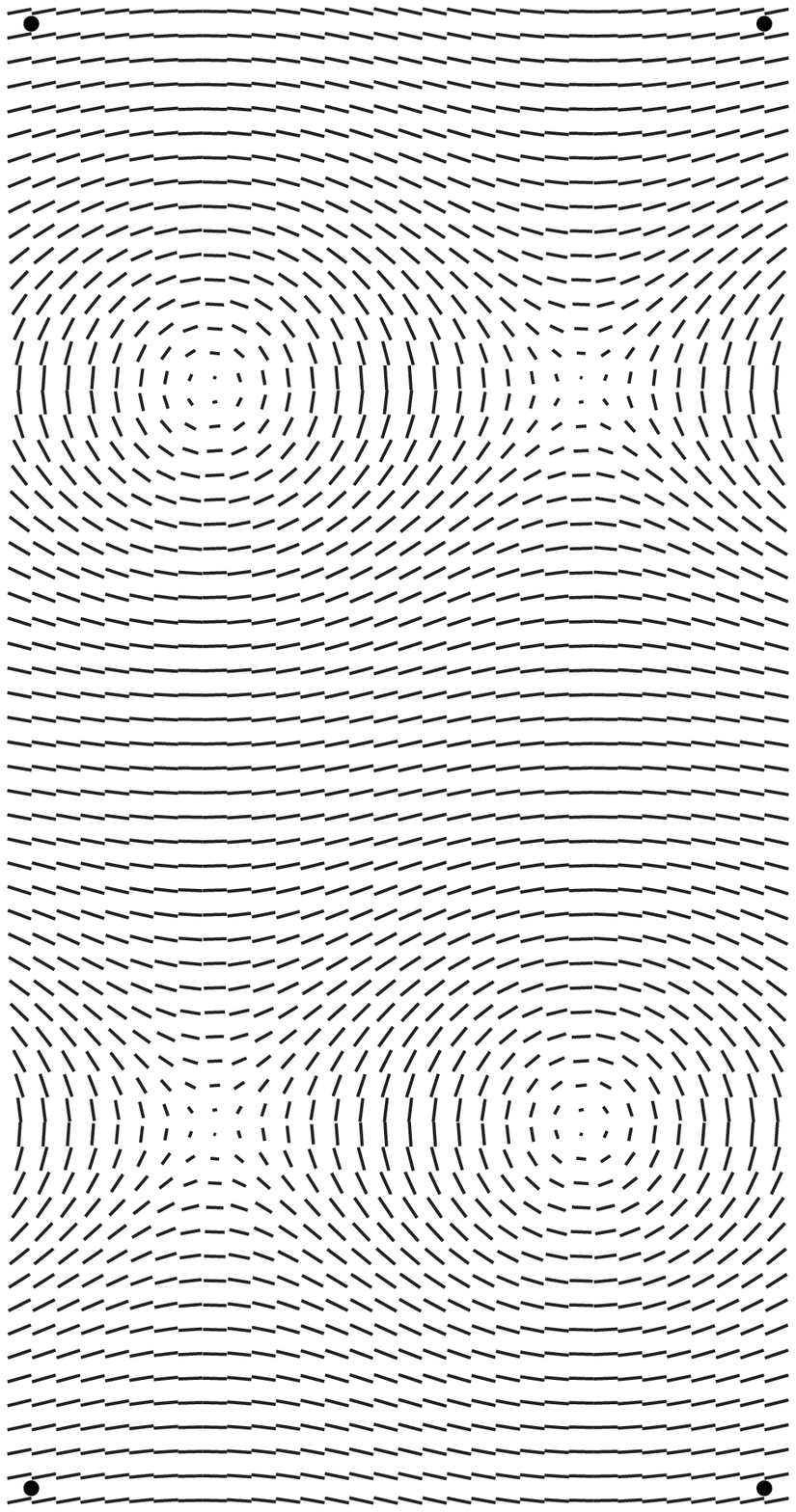}
\bigskip
\caption{ 
Locked vortex 3 (LV3) for $H=0.3H_{\rm d}$ and $\Omega
= 0.006 \Omega_{\rm d}$. The notation is the same as in Fig.\
\protect\ref{f.LV1}. The LV2 has the primitive rectangular Bravais
lattice and the space group $Pb'a'n$. The spacing of the plotted
$\hat{\bf l}$ vectors is $1.08\xi_{\rm d}$.
}\label{f.LV3}\end{center}\end{figure}

The LV3 structure can be understood as a modification of the LV1
structure. When the cores of the Mermin-Ho vortices shrink with
increasing magnetic field, there remains a large bending of $\hat {\bf
d}\approx\hat {\bf l}$ outside of the cores. The gradient energy in
this region can be reduced by rearranging the Mermin-Ho vortices. In
LV3 the Mermin-Ho vortices form infinite chains, as visible in Fig.\
\ref{f.LV3}.
The chains, or sheets consist of alternating circular and hyperbolic
units. Outside the sheets, the $\hat {\bf d}\approx\hat {\bf l}$ fields
are nearly constant and parallel (or antiparallel) to $x$. 

In contrast to all other vortex transitions, the change between LV1 and
LV3 seems to be of the second order. The lattice ratio $b/a$ grows
continuously from unity at the transition. At constant $H$, the $b/a$
ratio grows with decreasing $\Omega$ and reaches $3.4$ at our lowest
$\Omega=0.001$ at $H=0.6$.  We fit
$F=\Omega(3.62+0.37\Omega^{-0.36})$ at $H=0.3$ in the range
$\Omega=0.001\ldots 0.008$. Here all the three parameters are free in
the fit. 

It was demonstrated by Fujita and Ohmi that the LV1 is unstable to a
deformation \cite{FO96}. They found a structure where four Mermin-Ho
vortices form a unit that is separated by some distance from the other
units. We find that this structure has higher energy than the LV3.
  
\begin{figure}[bt]
\begin{center}\leavevmode
\includegraphics[width=0.8\linewidth]{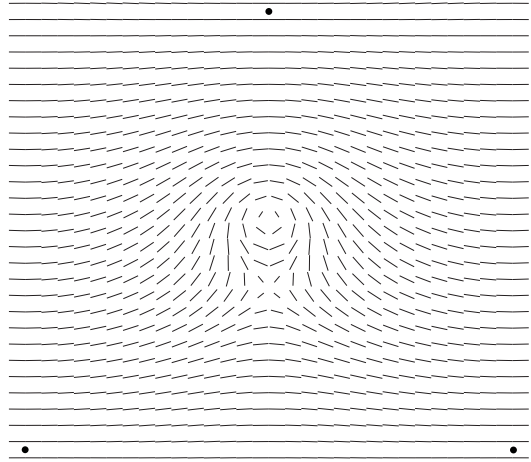}
\bigskip
\caption{ 
Locked vortex 2 (LV2) for $H=0.4H_{\rm d}$ and $\Omega
= 0.002 \Omega_{\rm d}$. The notation is the same as in Fig.\
\protect\ref{f.LV1}. The LV2 has the centered rectangular Bravais
lattice and the space group $C2'$. The spacing of the plotted $\hat{\bf
l}$ vectors is $1.97\xi_{\rm d}$. 
}\label{f.LV2}\end{center}\end{figure}

(iii) The locked vortex 2 (LV2) \cite{ZM85} is defined by $N=2$ and
$\nu_l=\nu_d=1$. The space group is $C12'1$, or shortly $C2'$, where
$C$ denotes the centering of the rectangular lattice. The LV2 represent
an alternative deformation of the LV1 when the cores of the elementary
vortices shrink in increasing magnetic field. Here the Mermin-Ho
vortices form pairs. Each pair consists of one circular and one
hyperbolic unit, see Fig.\ \ref{f.LV2}.
Outside such a pair, the $\hat {\bf d}\approx\hat {\bf l}$ fields are
nearly constant and parallel to $x$. This special direction breaks the
hexagonal lattice symmetry, which otherwise could apply to such well
separated vortices. Therefore, we expect the lattice structure is
centered rectangular. We find that the ratio of the two lattice
constants $b/a\approx 1.8$. This is less than $b/a=\sqrt{6}$, which is
obtained from the hexagonal lattice (having $b/a=\sqrt{3}$) by scaling
$x$ and $y$ according to the anisotropy of the superfluid density
($\rho_\perp=2\rho_\parallel$). 

The LV2 has only one point symmetry $2_y'$, which means a rotation by
$\pi$ around the $y$ axis combined with time inversion. In previous
$^3$He literature the $2'$ symmetry was denoted by $w$ \cite{SVrev}.
The symmetry implies that $\hat d_y$ and $\hat l_y$ change signs when
$x\rightarrow -x$ in Fig.\ \ref{f.LV2}, and other components of
$\hat{\bf d}$ and $\hat{\bf l}$ remain unchanged. The symmetry
transformation of $\hat {\bf m}$ and $\hat {\bf n}$ depends on the
specific choice of the phase factors.

The LV2 is doubly degenerate. The degenerate forms are obtained from
each other by exchanging the positions of the circular and hyperbolic
Mermin-Ho vortices. Formally this can be done by the operation $m_y'$.
In the calculation we have only studied the simplest case where all LV2
vortices have the same orientation. If both degenerate forms are
present simultaneously, it would lead to a larger primitive cell or, in
an extreme case, absence of periodicity. 

We find the energy $F=\Omega[-1.14+1.37\ln(1/\Omega+1/0.00412)]$ at
$H=0.3$ in the range $\Omega=0.001\ldots 0.008$. In this fit we kept
the constant 1.37 multiplying the logarithm fixed. This
constant arises from the flow far from the vortex cores, and it should
become exact in the limit $\Omega\rightarrow 0$. The numerical value
1.37 is calculated in Ref.\ \onlinecite{VH81}. The constant -1.14 can be
interpreted as the core energy of the vortex line. The constant 0.00412
describes the angular velocity where the cores of the vortices start to
overlap. In the range of our calculation the LV2 has higher energy than
LV3. However, it is expected that the LV2 becomes absolutely stable in
the limit $\Omega\rightarrow 0$. Extrapolating the expressions of the
energies we can estimate that this takes place at $\Omega\lesssim
0.0005$.

\begin{figure}[bt]
\begin{center}\leavevmode
\includegraphics[width=0.8\linewidth]{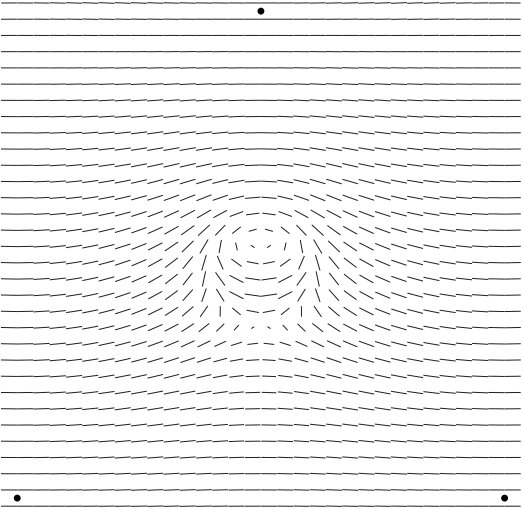}
\bigskip
\caption{ 
Continuous unlocked vortex (CUV) for $H=0.3H_{\rm d}$
and $\Omega = 0.02 \Omega_{\rm d}$. The notation is the same as in
Fig.\ \protect\ref{f.LV1}. Similar to LV2, the CUV has the centered
rectangular Bravais lattice and the space group $C2'$. The spacing of
plotted $\hat{\bf l}$ vectors is $0.59\xi_{\rm
d}$. 
}\label{f.CUV}\end{center}\end{figure}

(iv) The continuous unlocked vortex (CUV) has $N=2$, $\nu_l=1$, and
$\nu_d=0$
\cite{SV83,ZM84,VSF84,O84,explett,HKS85,MZ85a,MZ85b,F85,ZM85,F87,TPM92}.
The space group is $C2'$. The CUV is similar to the LV2 with respect
to $N$, $\nu_l$, and the space group, see Fig.\ \ref{f.CUV}. The
crucial difference compared to the LV2 is that $\nu_d$ vanishes.
Vortices where $\hat {\bf d}$ and $\hat {\bf l}$ differ essentially
from each other ($\nu_d\not =\nu_l$) are called ``unlocked''.

The principle in all unlocked vortices is to minimize the field energy
(\ref{e.fh}) in the first place, and therefore they are more economical
in large fields $H\gtrsim H_{\rm d}$ than the locked structures. In the
CUV, $\hat {\bf d}$ is approximately constant and parallel to $x$
everywhere. The region where $\hat {\bf d}$ and $\hat {\bf l}$ differ
from each other is called a ``soft core''. Its size $\sim \xi_{\rm d}$
is determined by the balance of the kinetic energy of ${\bf v}$
(\ref{e.grad}) outside of the soft core and the dipole energy
(\ref{e.fd}) inside. (Note that the total gradient energy of $\hat {\bf
l}$ in the core is approximately independent of the size.)

The first suggestion for CUV had the $m'$ symmetry \cite{SV83}. Similar
to several previous calculations, we find that the $2'$ symmetric form
has a lower free energy at all values of $\Omega$ and $H$
\cite{ZM84,explett,MZ85a,F87}. The $m'$ symmetric form seems to
correspond to a saddle point of the free energy.

Similar to the LV2, the CUV probably has the centered rectangular
lattice, and we find that $b/a\approx 2$.  The energy of the CUV
depends surprisingly much on the magnetic field. At $H=0$ we fit
$F=\Omega[-0.728+1.37\ln(1/\Omega+1/0.219)]$  but at
$H=0.6$ we find $F=\Omega[-0.591+1.37\ln(1/\Omega+1/0.0894)]$.
The expressions are most accurate in the range
$\Omega=0.008\ldots 0.03$. Similar to the LV2, the value 1.37 is kept
fixed in fitting the two other parameters. The interpretation of the
other parameters is the same as for LV2. The curving of the phase
boundary between CUV and VS (Fig.\ \ref{f.diagram}) arises from the
field dependence of CUV, as VS seems to be rather insensitive to $H$. 

According to our earlier calculation, the transition between the CUV
and the LV2 takes place at $H=0.4$ in the limit $\Omega\rightarrow 0$
\cite{TK94}. This agrees with the present calculation. We find the
triple point between the LV1, the LV3, and the CUV at $H=0.29$ and
$\Omega=0.014$. 

\begin{figure}[bt]
\begin{center}\leavevmode
\includegraphics[width=0.7\linewidth]{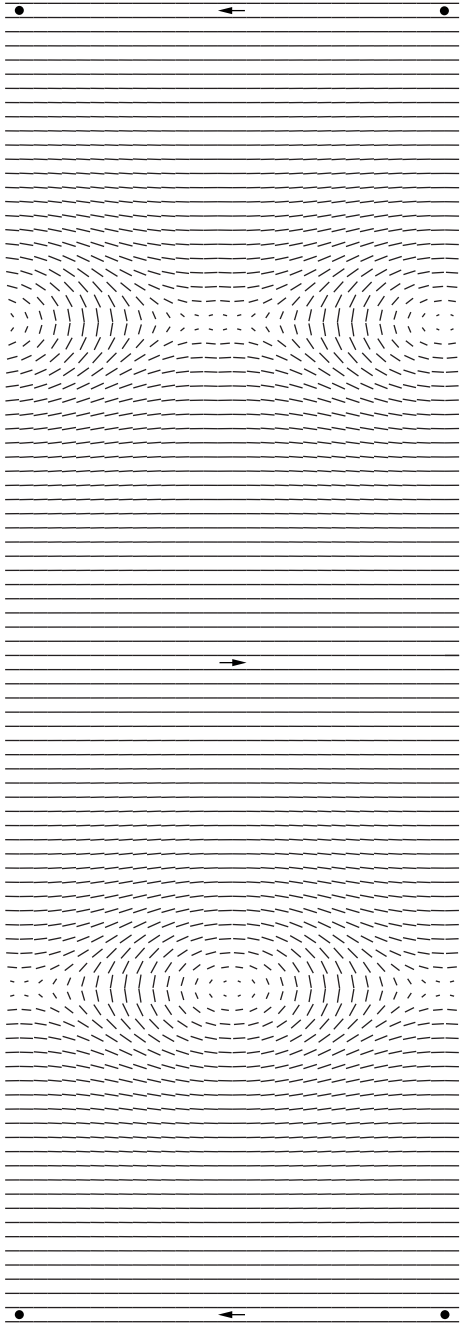}
\bigskip
\caption{ 
Vortex sheet (VS) for $H=0.2H_{\rm d}$ and $\Omega =
0.03 \Omega_{\rm d}$. The notation is the same as in  Fig.\
\protect\ref{f.LV1}. The VS has the primitive rectangular Bravais
lattice and the space group $Pb'a'n$. The primitive cell contains two
vortex sheets parallel to $x$. The arrows denote the opposite
directions of $\hat{\bf l}$ on different sides of the sheets. The
spacing of plotted $\hat{\bf l}$ vectors is $0.39\xi_{\rm d}$. 
}\label{f.VS}\end{center}\end{figure}

(v) The vortex sheet (VS) has $N=4$, $\nu_l=2$, and $\nu_d=0$
\cite{VS,VSintro,VSexp}. The likely space group is $Pb'a'n$, similar
to the LV3. Here $N$, $\nu_l$, and $\nu_d$ are all the same as for the
CUV except multiplication by 2. The crucial difference between the VS
and the CUV becomes evident when their primitive cells are stacked one
after another: in the CUV the soft cores form a two-dimensional lattice
of lines while in the VS they form a series of equidistant planes
parallel to $x$, see Fig.\ \ref{f.VS}. The $\hat{\bf d}$ vector is
approximately constant and parallel to $\hat{\bf x}$ everywhere.

The close similarity of the CUV and the VS is illuminated if one thinks
bending a VS and closing it to a cylinder. The CUV represents the
smallest among such cylinders because it contains just one periodic unit
of a vortex sheet. This relationship is also evident in the ansatz
forms given in the Appendix. 

We have  discussed above how the LV1 is transformed to the LV2 by
pairing the Mermin-Ho units. Then the CUV evolved from the LV2 when the
dipole locking was removed. After that the VS was developed by opening
the cylindrical structure of the CUV. We can now return to the starting
point by noting that forcing dipole locking in the VS gives the
structure of LV1 and LV3. The close similarity of the VS and the LV3 is
evident from the symmetry groups and from Figs.\ \ref{f.LV1} and
\ref{f.LV3}. 

The $\hat{\bf l}$ vector in the VS is approximately parallel to $x$
outside of the soft cores, but it has opposite directions on the two
sides of the sheet. This implies that the primitive cell must contain
(at least) two neighboring sheets. This is the reason for the double
size of the primitive cell compared to the simplest lattice structure of
the CUV. The surface tension of the sheet makes the primitive cell
rather short in the $x$ direction. We note that the present definition
of $b$ according to the  primitive rectangular Bravais lattice is twice
as large as in Refs.\ \onlinecite{VSintro,VS}, and
\onlinecite{VSexp},  where $b$ denotes the distance between neighboring
sheets. Because the sheets are far from each other, the sliding of the
sheets relative to each other probably leads to a negligible change in
the energy.  

The properties of the VS depend very weakly on the magnetic field in
the studied region $H\le 0.6$. The results of our calculations can be
represented by $b/a = 1.26\Omega^{-0.255}$ and
$F=1.35\Omega^{0.664}$ for
$\Omega = 0.016\ldots 0.07$. 
Both these quantities are slightly larger than obtained from the
twist-section model\cite{VSintro}, which gives 
$b/a={1\over\pi}(18/\Omega)^{1/3}$ and
$F={1\over 2}(18\Omega^2)^{1/3}$. The uniform winding model is a
slightly more complicated variational ansatz\cite{VSintro,VSexp}. It
gives an upper bound for the energy that is 10\% higher at 
$\Omega = 0.016$ and 3\% higher at $\Omega = 0.07$.
In the limit $H\gg 1$ we find the transition between CUV and VS at
$\Omega = 0.022$.

A new feature in the phase diagram is  that both the CUV and the
VS are stable also in zero field. 
Although it has not been stated explicitly, the crossing of the
energies of the LV1 and the CUV at $H=0$ appears also in previous
literature. Comparison of the energies given in Refs.\
\onlinecite{FNOT78} and \onlinecite{F87}, for example, gives it at
$\Omega=0.023$. It was calculated in Ref. \onlinecite{FSS83} 
that there is a transition from the LV1 to a singular
$\hat{z}$ vortex when $\Omega=0.11$. This prediction has to be revised
because both the CUV and the VS have a much lower energy at this
rotation velocity. If there is a transition to the $\hat{z}$ vortex, it
takes place at a much higher $\Omega$ than expected in Ref.
\onlinecite{FSS83}. 

The energy difference of the CUV and the VS relative to the LV1 at
$H=0$ arises from competition of the dipole energy and the gradient
energy of $\hat{\bf d}$. The LV1 is stable at a low $\Omega$ because
the dipole energy is minimized in the first place, and the gradient
energy associated with $\hat{\bf d}$ is not important. With increasing
$\Omega$ the gradient energy becomes larger. At the transition point,
it becomes more economic to arrange $\hat{\bf d}$ approximately
constant although it means increased dipole energy in the soft core of
the CUV or the VS. Based on purely dimensional considerations, this
transition was expected at $\Omega\sim 1$ ($\Omega\sim\Omega_{\rm
d}\approx 120$ rad/s in real units) \cite{FNOT78}. However, the present
calculation gives the transition between the CUV and the LV1 at
$\Omega=0.019$, which is almost two orders of
magnitude smaller than the naive expectation.   

(vi) The singular vortex (SV) has $N=1$ and the space group $C1m'1$, or
shortly $Cm'$ \cite{SV83,VSF84,F87}. No other vortex considered here
has $N=1$ because it is not possible for the continuous vortices as a
result of Eq.\ (\ref{e.contn}). The $\Omega$ dependence of the energy
can be written $F=\Omega[c(H,T)-0.70\ln\Omega]$ at small $\Omega$. Here
the factor $0.70$ arises from the flow field far from the vortex line
\cite{VH81}. This factor is approximately one half of the value for a
$N=2$ vortex line. Therefore, the singular vortex is favored over the
CUV at a low $\Omega$. 

Based on topological arguments alone, the structure of the singular
vortex could be very simple. For example, $\hat{\bf m}+{\rm i}\hat{\bf
n}=\exp({\rm i}\phi)(\hat{\bf y}+{\rm i}\hat{\bf z})$ and $\hat{\bf
d}=\hat{\bf l}$ everywhere except at the singular ``hard core'', where
these quantities are not defined. Here $\phi$ is the azimuthal angle.
However, energetics prefers a more complicated structure that has  a
soft core in addition to the hard core \cite{SV83}. This is because a
structure with radial $\hat{\bf l}$ and constant $\hat{\bf m}=\hat{\bf
z}$ has lower energy than the simple vortex. Outside of the soft core,
$\hat{\bf d}$ and $\hat{\bf l}$ are both nearly parallel to $x$.
Similar to the LV2 and the CUV, this gives rise to a centered
rectangular lattice.
 
The hydrodynamic approximation used for the calculation of the
continuous vortices is insufficient in the hard core of the SV.
Therefore we have not calculated the function $c(H,T)$ in the energy
$F$. Contrary to the continuous vortices, there is also a temperature
dependence $c(H,T)\approx c(H)+0.7\ln(1-T/T_{\rm c})$ because the size
of the hard core depends on $T$. In order to get an idea of the
complete phase diagram, the phase boundary of the SV is included in
Fig.\ \ref{f.diagram} (dashed lines) by two arbitrarily chosen constant
values of $c(H,T)$. 

The core structure of the SV can in principle be calculated using the
Ginzburg-Landau theory, but there are two difficulties. Firstly, this
theory introduces additional parameters (such as the coefficients of
the five energy terms that are of the fourth order in the order
parameter matrix) whose values are not well known. So the accuracy of
the results would be less than in the hydrodynamic theory of continuous
vortices. Secondly, the numerical calculation is difficult because the 
length scales associated with the soft ($\xi_{\rm d}\approx 10\ \mu$m)
and hard cores ($\approx 10\ $nm) are very different. 

\begin{figure}[bt]
\begin{center}\leavevmode
\includegraphics[width=0.8\linewidth]{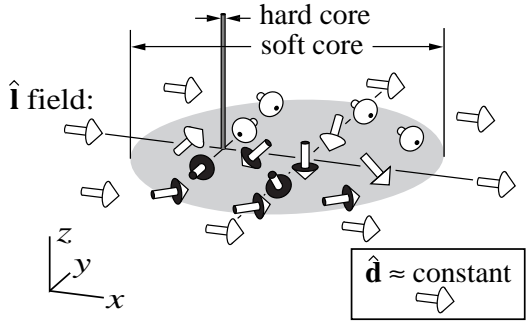}
\bigskip
\caption{ 
Sketch of the core of the singular vortex (SV). The
arrows denote $\hat{\bf l}$ and $\hat{\bf d}$.
The soft core, where $\hat{\bf d}$ and $\hat{\bf l}$ differ
considerably, appears as shaded. 
}\label{f.SV}\end{center}\end{figure}

We have made numerical simulations with Ginzburg-Landau theory where the
difference in the soft and hard-core scales is arbitrarily reduced. We
cannot expect any quantitative results from such a calculation, but we
believe that the following qualitative results are valid independently
of our approximation. The structure around the hard core is  
\begin{equation} \pm\hat{\bf l}=\hat{\bf y}\sin\phi+\cos\phi(\hat{\bf
x}\cos\eta+\hat{\bf z}\sin\eta), \label{e.sing} \end{equation}  where
$\eta$ is a constant angle, see Fig.\ \ref{f.SV} for illustration. The
original suggestion \cite{SV83} has $\eta=\pi/2$ whereas the minimum
energy of the structure (\ref{e.sing}) prefers $\eta=0$ \cite{SVrev}.
The true structure is likely to fall between these limits because
$\eta=0$ would imply a large gradient energy (\ref{e.grad}) in matching
the hard core (\ref{e.sing}) with the constant $\hat{\bf l}$ outside of
the soft core.
 
There is no circulation around the hard core. Thus all the circulation
arises from the soft core, which is qualitatively described by the
numbers $\nu_l=1/2$ and $\nu_d= 0$. The centers of the hard and soft
cores are displaced from each other in the $x$ direction. The vortex
has symmetry $m_y'$, which in previous $^3$He literature was called
$v$. In terms of the vectors this means that $d_y$ and $l_y$ change
signs when $y\rightarrow -y$, and other components of $\hat{\bf d}$ and
$\hat{\bf l}$ remain unchanged. The SV is doubly degenerate. The two
forms are obtained from each other by the symmetry operation $2_y'$.

All the discussion above was for the case where the field ${\bf H}$
is parallel to $\bbox{\Omega}$. As far as we know, other directions are
considered only for the CUV \cite{MZ85b,O84} and the VS \cite{VSexp}.
Generally, it can be expected that the effect of field direction is not
large in unlocked vortices, where $\hat{\bf d}$ is approximately
constant, and thus $f_{\rm h}$ (\ref{e.fh}) can equally be minimized
for arbitrary direction of ${\bf H}$.   

\section{Comparison with experiments}\label{s.exp}

In order to compare the calculated phase diagram with measurements, one
needs to know how to prepare the equilibrium state in experiments. This
is not simple because the energy barriers separating the different
vortex types are generally so large that it is difficult to induce
any transitions \cite{diag}. So the decay of a metastable vortex type to
the equilibrium type may be so slow that it cannot be observed. Also,
if the rotation is started in the superfluid state, the vortex type
that nucleates is generally not the equilibrium one. For example,
only continuous vortices are nucleated if the rotation is started in
the superfluid state; no singular vortex has been observed by this
method. 

The only exception to the above seems to be the region very near the
superfluid transition temperature $T_{\rm c}$. There the energy
barriers separating the different vortex types are smallest. A
practical way to perform the experiment is to cool slowly from the
normal state ($T>T_{\rm c}$) to the superfluid state at constant
$\Omega$ and $H$. This procedure is expected to yield a state near the
equilibrium one. It is important to remember, however, that the details
of the transition in the presence of a thermal gradient and a magnetic
field may be rather complicated \cite{diag}.  Another limitation of the
experiments is that they do not resolve the difference between the
three types of the LV.    

Also needed for the comparison are the values of $\Omega_{\rm d}$ and
$H_{\rm d}$. We estimate $\Omega_{\rm d}\approx 120$ rad/s, and $H_{\rm
d}\approx 2.0$ mT at 29 bar pressure. These are based on a
weak-coupling analysis corrected by the enhancement of the energy gap
according to Ref.\ \onlinecite{SRrev}. In addition, we have used the
measured shift of the transverse NMR resonance frequency in the A
\cite{Osheroff} or the B phase \cite{AKP}, both of which give
essentially the same result. The gap enhancement, which in Ref.\
\onlinecite{SRrev} is given for the B phase as a function of the 
specific heat jump $\Delta C_{\rm B}$, is applied to the A phase by
replacing $\Delta C_{\rm B}$ by ${6\over 5}\Delta C_{\rm A}$.

The comparison of the experimental and theoretical phase diagrams is
shown in Fig.\ 3 of Ref.\ \onlinecite{diag}. The qualitative agreement
is good. A slight difference is that the vortex sheet is not observed
in the experiments at the maximal angular velocity 3 rad/s although
according to the present calculation it should show up above 2.6 rad/s,
assuming $\Omega_{\rm d}=120$ rad/s.  Possible explanations are that
$\Omega_{\rm d}$ is larger than we estimated or the cooling
through $T_{\rm c}$ does not accurately produce the equilibrium state.

The six vortex types discussed above seem to be able to explain all
the experiments that have been made in rotating bulk $^3$He-A. We
comment here on one controversial experiment. Torizuka {\it et al.}
\cite{TPMV91} observe a transition in the rotating state at $\Omega=3$
rad/s when the rotation velocity was varied at constant $H=0$. The
original interpretation in the same reference postulated a layer
of vortices on the container wall. We consider this interpretation
unlikely because such a layer is probably unstable, as pointed out in
Ref.\ \onlinecite{PKKKT}. The present calculation now gives the
possibility that the transition could be from the LV1 to either the CUV
or the VS. Unfortunately, the collected experimental data does not seem
sufficient to identify the structure at large $\Omega$
\cite{PKKKT,TPM92}. 

\section{Conclusion}\label{s.conclusion}

We have presented numerical calculations of the vortex structures in the
Ginzburg-Landau region, and constructed a phase diagram in the
$H-\Omega$ plane. There is in principle no difficulty in
extending these  calculations to lower temperatures. Although the phase
diagram is not accessible experimentally at low temperatures, the
calculation of the vortex structure would form the basis for a
calculation of the NMR frequency shifts, which have been measured
accurately. 

Our search of vortex types was based on previous suggestions. There may
well be structures which could not have evolved from the initial
guesses we have used. In particular, only the simplest periodic
structures were tested. From the experimental point of view, it seems
that there is at present no need to introduce new types of vortices
that are stable in bulk $^3$He-A. That may change, however, when new
regions are studied and more accurate measurements are done. In
particular, low temperatures, high rotation velocities, high magnetic
fields, the neighborhood of the A$_1$ phase, and restricted geometries
could be studied. The studies could also be extended to metastable
structures. For example, the LV was identified from its metastable
modification in high field, which more appropriately should be
classified as a new type of vortex \cite{diag}.

\section*{Acknowledgments}

We wish to thank A. Borovik-Romanov, P. Hakonen, M. Krusius, O.
Nevanlinna, B. Pla\c{c}ais, J. Simola, E. Sonin, and G. Volovik for
useful discussions and R. H\"anninen for help in calculations.

\section*{Appendix}\label{s.appendix}

We give approximate expressions for the order parameter in different
vortex structures. These can be used as initial guesses to produce the
stable vortices discussed in Section \ref{s.results}.

All structures can be simply represented using Euler angles ($\alpha$,
$\beta$, $\gamma$) but choosing $x$ as the polar direction:  
\begin{eqnarray} \hat{\bf l}&=&\hat{\bf x}\cos\beta+\sin\beta (\hat{\bf
y}\cos\alpha+\hat{\bf z}\sin\alpha)\label{e.l} \\ \hat{\bf
m}+{\rm i}\hat{\bf n}&=&[-\hat{\bf x}\sin\beta+\cos\beta (\hat{\bf
y}\cos\alpha+\hat{\bf z}\sin\alpha)\nonumber \\  &&+{\rm i}(-\hat{\bf
y}\sin\alpha+\hat{\bf z}\cos\alpha)]\exp(-{\rm i}\gamma).
\label{e.mn}\end{eqnarray} In these coordinates the superfluid velocity
(\ref{e.vel}) is  \begin{equation}
\tilde{\bf v}_{\rm s}=-\bbox{\nabla}\gamma-\cos\beta\bbox{\nabla}\alpha.
\label{e.velE} \end{equation} Depending on the vortex type, we present
the Euler angles as functions of either rectangular ($x$, $y$, $z$) or
cylindrical coordinates ($r$, $\phi$, $z$). Unless explicitly stated
otherwise, the ansatz forms have the same symmetries as the vortices
that they represent.

The VS has $\hat{\bf d}=\hat{\bf x}$ and $\alpha=-\gamma=\pi/2-2\pi x
\mathop{\rm sgn}(y)/a$, where $\mathop{\rm sgn}(y)$ denotes the sign of 
$y$. $\beta$ is a monotonic function of $y$ so that $\beta(-b/2)=-\pi$, 
$\beta(0)=0$, and $\beta(b/2)=\pi$. Especially at low velocities this
function is strongly nonlinear so that all the change of $\beta$ takes
place in narrow regions (thickness $\sim 1$) at the two vortex sheets,
which are located at $y=\pm b/4$. This form of the order parameter is
for a gauge where ${\bf v}_{\rm n}=-2\Omega y\hat{\bf x}$. The
transformation to the more usual gauge ${\bf v}_{\rm n}=\bbox{\Omega}
\times{\bf r}$ is obtained by including an extra factor $\exp({\rm
i}\Omega xy)$ multiplying the right hand side of Eq.\ (\ref{e.mn}). The
lattice constants are constrained by $ab=4\pi/\Omega$ (\ref{e.area}).

An approximation to LV1 and LV3 is the same as for VS except
that $\hat{\bf d}=\hat{\bf l}$. The best guess for LV1 corresponds to
$a=b$ and nearly linear $\beta(y)$. In spite of these choices, this
ansatz has less symmetry than the converged solution for LV1. A
symmetric but more complicated ansatz was suggested in Ref.\
\onlinecite{FNOT78}. A third alternative is to glue together ansatzes
of four Mermin-Ho vortices. 

The CUV has $\hat{\bf d}=\hat{\bf x}$ and $\alpha=-\gamma=\phi$.
$\beta$ is a monotonic function of the radius $r$, which has
$\beta(0)=0$ and $\beta(r)\approx\pi$ for $r>r_0$. Here the radius
$r_0$ is of the order of a few units of length ($\xi_{\rm d}$). 

The LV2 is similar to CUV except that $\hat{\bf d}=\hat{\bf l}$. $r_0$
is of the order of $H^{-1}$ in dimensionless units.

The SV can be generated by $\hat{\bf d}=\hat{\bf x}$, $\gamma=0$, and
$\alpha=\pi/2+\phi$. $\beta$ is a monotonic function of $r$, which has
$\beta(0)=\pi/2$ and $\beta(r)\approx\pi$ for $r>r_0$. Here $r_0$ is
approximately unity. This ansatz form has more symmetry than is present
in a converged solution. 


\begin{thebibliography}{99}
\bibitem{tinote} This mapping was used without explicit mention in Ref.
\onlinecite{T87} for classification of vortices in $^3$He-B.
\end{thebibliography}
 \end{document}